# A Machine Learning Approach to Predicting Pore Pressure Response in Liquefiable Sands under Cyclic Loading

Yongjin Choi, M.S., S.M.ASCE[1]; Krishna Kumar, Ph.D., Aff. M. ASCE[2].

[1]Department of Civil, Architectural and Environmental Engineering, The University of Texas at Austin, TX, E-mail: yj.choi@utexs.edu
[2]Department of Civil, Architectural and Environmental Engineering, The University of Texas at Austin, TX, E-mail: krishnak@utexas.edu

## ABSTRACT

Shear stress history controls the pore pressure response in liquefiable soils. The excess pore pressure does not increase under cyclic loading when shear stress amplitude is lower than the peak prior amplitude -- the shielding effect. Many sophisticated constitutive models fail to capture the shielding effect observed in the cyclic liquefaction experiments. We develop a data-driven machine learning model based on the LSTM neural network to capture the liquefaction response of soils under cyclic loading. The LSTM model is trained on 12 laboratory cyclic simple shear tests on Nevada sand in loose and dense conditions subjected to different cyclic simple shear loading conditions. The LSTM model features include the relative density of soil and the previous stress history to predict the pore water pressure response. The LSTM model successfully replicates the pore pressure response for three cyclic simple test results considering the shielding and density effects.

## INTRODUCTION

During an earthquake, the pore pressure response of saturated soils controls their behavior and is related to liquefaction triggering. Liquefaction occurs when excess pore water pressure exceeds effective stress, resulting in a significant loss of strength and stiffness.

Cyclic simple shear tests on saturated Nevada sand under different densities and input stresses (Sideras, 2019) show that the stress history plays a vital role in the generation of pore water pressure. When the soil experiences a shear stress pulse amplitude larger than the prior peak amplitude of the stress history, it causes a rapid increment in pore pressure. On the other hand, when the stress amplitude is lower than the prior peak amplitude, the stress does not contribute to any pore pressure generation unless the soil reaches the phase transformation state. This stress history-dependent response of pore water pressure is called the shielding effect.

Ishibashi et al. (1977) proposed a relationship to predict pore water pressure build-up of saturated sandy soil with loading cycles based on the number of cycles, shear stress at the previous stress, and pore water pressure at the previous cycle. This predictive model can account

– 1 –

for the discrete pore water pressure value after each cycle of harmonic loading but not the continuous build-up of stresses with time. Because earthquake loadings are irregular, not harmonic, and the pore water pressure responds to the continuous stress history for every time step, not after each cycle of previous stress, we should investigate the behavior in a continuous framework.

Sideras (2019) investigated if numerical models could capture the continuous response of pore water pressure, including the shielding effect (Fig. 1). Sideras used the Pressure Dependent Multi-yield 02 (PDMY02) (Yang et al., 2008) and the PM4Sand (Boulanger and Ziotopoulou, 2017) model to simulate single-element cyclic simple shear tests on OpenSees and FLAC. Both PDMY and PM4Sand models could not capture the shielding effect observed in the laboratory experiments. Fig. 1 shows the pore pressure response for cyclic simple shear test on Nevada sand. The pore water pressure $r_u$ is the ratio between excess pore water pressure and effective stress. Laboratory tests and constitutive models predict the increase in the pore pressure ratio $r_u$ until around 36 seconds as shear stress ($\tau$) amplitude grows (region A). At 36 seconds, the sample experiences a large stress pulse, and the subsequent stress does not exceed its amplitude until 57 seconds. As a result, $r_u$ in test data maintains a nearly constant level. However, both constitutive models show a clear increase in $r_u$ during this period (region B). From 57 to 60 seconds, $r_u$ begins to increase again in the test data because the loading amplitudes exceed the prior one (region C). The constitutive models also show these increases but to a lower extent. Region D again experiences the shielding effect; unlike previous times, the models capture the shielding effect in this region. In addition, the model predictions of time to liquefaction (when $r_u$ reaches 1.0) slightly deviate compared to the test data. The constitutive models also do not reproduce the fluctuations in the pore pressure response.

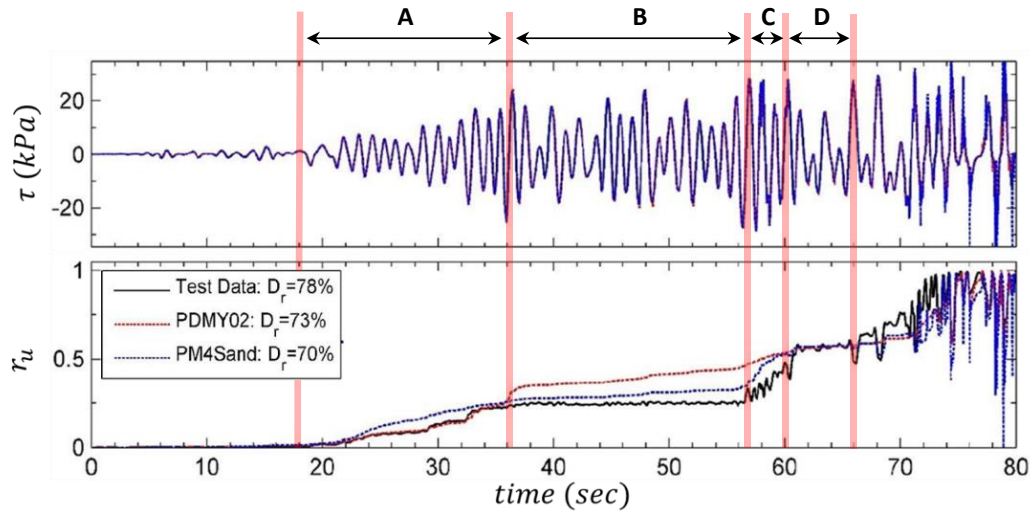

**Figure 1. Pore pressure response $r_u$ to cyclic simple shear test using PDMY02 and PM4Sand models (Siders, 2019), where τ is the input shear stress.**

Recently, there have been efforts in the geotechnical field to describe soil behaviors using a data-driven approach. Given a sufficient amount of data on the behavior of pore water pressure responding to various input stresses, one can use machine learning and neural networks to find a model that best explains such behavior. We use a long short-term memory (LSTM) neural network (Hochreiter and Schmidhuber, 1997) trained on cyclic simple shear test data from (Kwan et al., 2017) to capture the pore pressure response.

**METHOD**

**Long short-term memory neural network.** Long Short-Term Memory (LSTM) network is a variant of the recurrent neural network (RNN). RNN can process and predict sequential data and is capable of remembering patterns observed in the input sequence. However, RNN fails to model long sequences as it suffers from vanishing and exploding gradient problems (Bengio et al., 1994). LSTM solves this problem by introducing forget, input, and output gates (Fig. 2). These gates, which are a 0-1 valued matrix, regulate what information from the previous sequence and current feature to consider or forget, and which information to pass to the next cell. The previous information is represented as the previous cell state ($C_{t-1}$) and hidden output ($h_{t-1}$). The current feature is $x_t$. The hidden output of the current cell is $h_t$ and the current cell state is $C_t$.

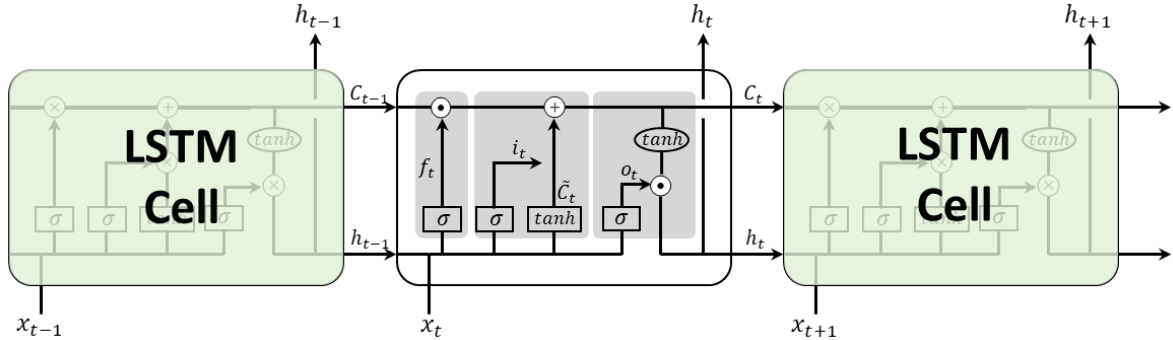

**Figure 2. LSTM cell architecture. The subscript $t$ stands for the time step. $x$, $h$, and $c$ are the input, hidden output, and cell state. $f_t$, $i_t$, and $o_t$ are the forget, input, and output gate at $t$. $\sigma$ and $tanh$ are the sigmoid and hyperbolic tangent function. $\odot$ and $\oplus$ are the element-wise product and summation.**

The details of the LSTM cell are as follows. Forget gate ($f_t$) regulates how much information of the previous cell state ($C_{t-1}$) to bring into the current cell state by referring to the previous hidden output ($h_{t-1}$) and current feature ($x_t$) (Eq. 1). $C_{t-1}$ is multiplied by $f_t$ (Eq. 5) squeezing values of elements in $C_{t-1}$ to be smaller or make it zero if they should not be included in the current cell state. The other gates do a similar operation. Input gate ($i_t$) regulates the extent of candidates ($\widetilde{C}_t$) (Eq. 3) of the new information ($x_t$) to be included in the current cell state (Eq. 2). The result of the input gate is added to the cell state determined in the forget gate to make the final current cell state ($C_t$) (Eq. 5), which will be passed to the next cell. Output gate ($o_t$) regulates which output to transfer to the next cell ($h_t$) based on $C_t$, $x_t$, and $h_{t-1}$ (Eq. 4).

$$f_t = \sigma(W_{xf}x_t + W_{hf}h_{t-1} + b_f) \tag{1}$$

$$i_t = \sigma(W_{xi}x_t + W_{hi}h_{t-1} + b_i) \tag{2}$$

$$\widetilde{C}_t = tanh(W_{xc} \times x_t + W_{hc} \times h_{t-1} + b_c) \tag{3}$$

$$o_t = \sigma(W_{xo}x_t + W_{ho}h_{t-1} + b_o) \tag{4}$$

$$C_t = f_t \odot C_{t-1} + i_t \odot \widetilde{C}_t \tag{5}$$

$$h_t = o_t \odot tanh(C_t) \tag{6}$$

$W$s and $b$s are weights and biases for corresponding gate operations. The first subscripts, $x$ and $h$, denote input and hidden output. The second subscripts, $f$, $i$, $o$, and $c$, denote forget, input, output, and candidate.

Due to the advantage of this gate structure of the LSTM, it can efficiently remember and process the long sequential information that it has seen in order to find meaningful rules related to its output, which is the thing that RNN fails to do. Several studies successfully predicted the history-dependent mechanical response of materials or systems using LSTM. Zhang et al. (2019) used the LSTM for nonlinear structural seismic response prediction and found that the model is reliable and computationally efficient for nonlinear structural response prediction. Zhang et al. (2020) developed an LSTM model that can estimate both drained and undrained behaviors of sands under cyclic loadings. Zhang et al. (2021) applied the LSTM for modeling the static stress-strain behavior of saturated soils, and they found that the model is able to reproduce the behavior from both numerical and laboratory experiment data. Benabou (2021) designed an LSTM network that can properly capture the complex behavior of a viscoplastic alloy subjected to strain rate jumps, temperature changes, or loading-unloading cycles.

LSTM-based network model can also be an effective approach to find patterns of pore water pressure generation affected by previous shear stress history due to its inherent ability to take account of it. In this study, we designed and trained an LSTM-based neural network model to predict the pore water pressure based on information about previous shear stress that soil experiences and the relative density of the sample.

**Datasets.** We train our LSTM model on 12 laboratory cyclic simple shear test results (Kwan et al., 2017) on Nevada sand samples. Fig. 3 shows the input stresses and the corresponding pore pressure generated in samples prepared at different relative densities ($D_r$).

Dataset 1-8 are the test results from loose sand samples (Fig. 3a and b). Among them, the applied stresses for datasets 2-3 are modulated-up, which means the stress gradually increases, with different increasing rates, while the datasets 4-8 are modulated-down, which means the stress gradually decreases, with different rates as well. Note that dataset 1 is harmonic stress for baseline case. Datasets 9-12 are the results from dense sand samples with modulated-up stresses (Fig. 3c). The performance of our trained LSTM model is tested on the other 3 test results. These datasets include both modulated-up and down motions on loose and dense samples to check the versatility of the model. The shape of the motions and relative densities for these datasets are shown later in Fig. 6 and 7.

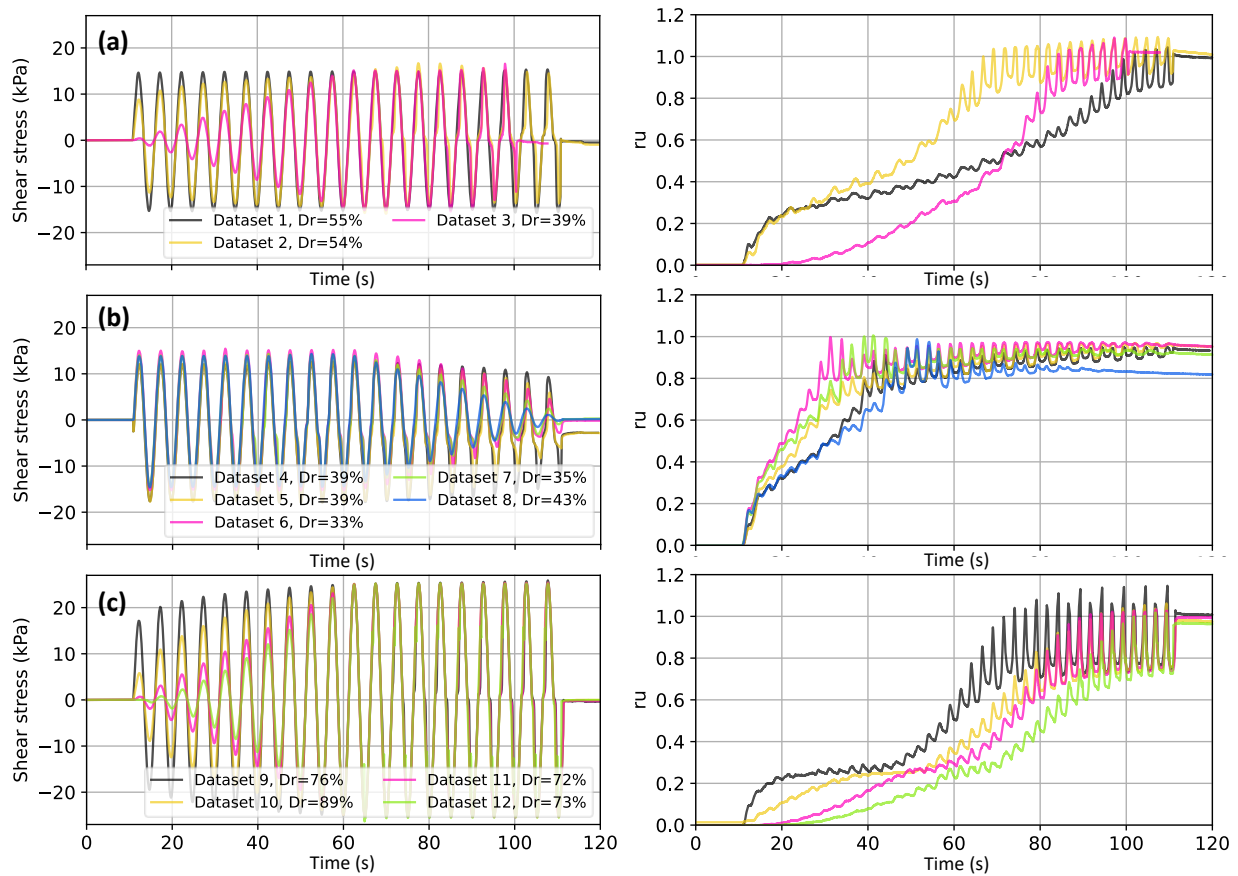

**Figure 3. The responses of pore water pressure ($r_u$) corresponding to different input shear stresses and sample relative densities ($D_r$) for 12 laboratory cyclic simple shear tests (Kwan et al., 2017): (a) one harmonic and two modulated-up stresses on loose samples; (b) five modulated-down stresses on loose samples; (c) four modulated-up stresses on dense samples.**

As explained in the introduction, the shielding effect describes the stress history-dependent behavior of pore water pressure generation; only a stress pulse that has a greater amplitude than the previous ones contributes to the pore water pressure build-up. This can be observed in the modulated-down stresses (Fig. 3b). $r_u$ shows a rapid rise in the early cycles because of stress upsurge. However, it stays still in the later part because the stress pulse keeps decreasing, which implies that $r_u$ generation is shielded by the prior peak stress amplitude. On the other hand, in the modulated-up motions (Fig 2a and c), $r_u$ gradually builds up at every loading cycle until it reaches 1.0 because the stress amplitude exceeds the prior one at every cycle. Obviously, the density of the sample also affects pore water pressure response. For the moduluate-up motions, the dense samples require larger stress amplitudes for reaching liquefaction (Fig. 3c) compared to loose samples (Fig. 3a).

**LSTM model structure and training.** To train our LSTM model, we sample training examples (feature-target pairs) from the 12 cyclic simple shear test datasets described in Fig. 3. The features ($\mathbf{X_t} = [T_t, \dots, T_{t+799}; \tau_t, \dots, \tau_{t+799}; D_{r,t}, \dots, D_{r,t+799}] \in \mathbb{R}^{3 \times 800}$) include the relative density ($D_r$) to consider the effect of density, shear stress ($\tau$) history of 800 time steps from $t$, which is about two loading cycles, for the model to refer to a few previous stress amplitudes, and time ($T$) itself that corresponds to each stress because the pore water pressure depends on the duration of excitation that soil sample is exposed. Note that we set the initial relative density constant for those 800 time steps. The target ($\mathbf{Y_t} = r_{u,t+800} \in \mathbb{R}^{1 \times 1}$) is the excess pore water pressure response of the subsequent time step.

The sliding window with the length of 800 samples training examples at every $t$ from each laboratory test data. Consequently, 103,538 examples are sampled. These are shuffled, and 80% of them are used for training the model, and 20% are used for the validation to determine the best model structure and to avoid overfitting to training samples.

Fig. 4 describes the determined model structure. The first LSTM layer takes a batch of input features. The three features at every time step ($x_k$) are passed to each of the LSTM cells with 128 units, and every cell outputs its own hidden output ( $y_k^{(1)}$ ). These outputs are passed to the second LSTM layer whose individual cell has 128 units, and the last cell outputs a hidden output ( $y_{t+799}^{(2)}$ ). This 128-dimensional array is processed by two consecutive fully connected (FC) layers and finally outputs $Y_t$ which is $r_{u,t+800}$.

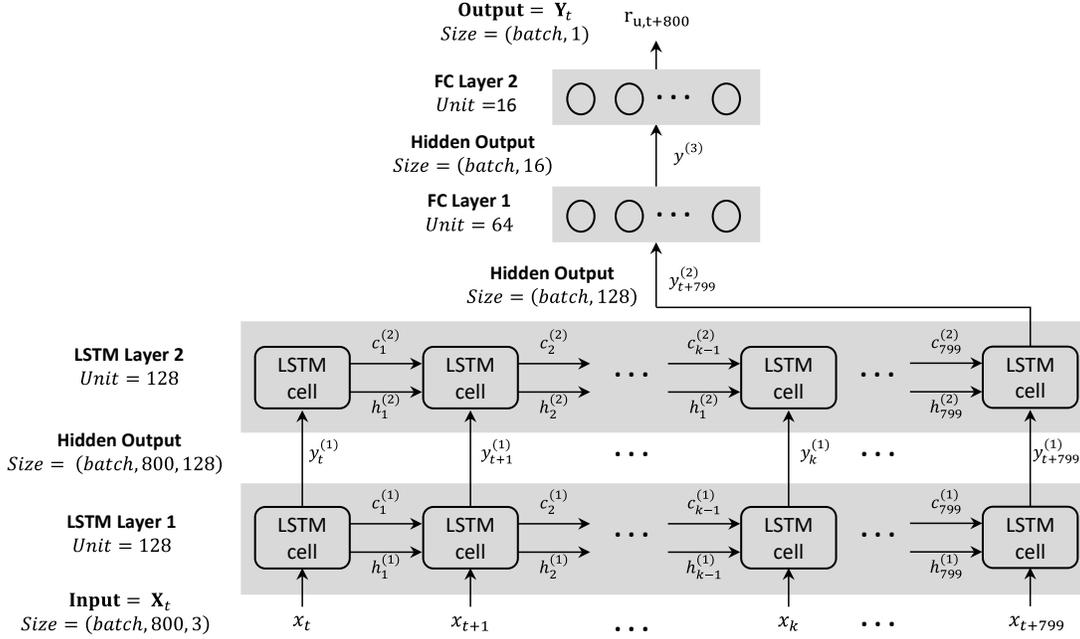

**Figure 4. The LSTM-based neural network model structure (modified from Zhang (2020)).
Note that $tanh$ activation functions are applied to all layers.**

All the units in the layers have their own parameters (weights and biases). An optimization algorithm finds the best values that minimize the difference between model output and the target value, which is called the loss, by seeing multiple batches of training examples. Every step the model sees a batch, the algorithm updates its current parameters to have a smaller loss. After the updating is done for the whole batch in the training example, the same process starts again. This iterative training step is called an epoch. We use the mean square error to compute the loss (Eq. 7) and the adaptive momentum estimation (Adam) optimization algorithm for minimizing it. Every training example is normalized to have values between 0-1 for improving the efficiency of the optimization process. Stress and times are normalized by the initial confining pressure and the final run time of the test, respectively.

$$loss = \frac{1}{n}\sum_{i=1}^{n}\left(Y_{t,predicted} - Y_{t,target}\right)^2 \qquad (7)$$

Where, $loss$ is the loss at a given set of model parameters; $n$ is the batch size, which is set to 32 in our case; $Y_{t,predicted}$ and $Y_{t,target}$ are predicted and target values correspond to each training example sampled at time $t$, respectively.

Fig. 5 shows the evolution of loss with epoch. After 100 epochs, the MSE for training and validation sets reaches $1.13 \times 10^{-4}$ and $8.93 \times 10^{-5}$. The fact that those values do not deviate much implies that there is no overfitting problem in our model. Therefore, no regularization or model complexity reduction are necessary. The total training took about 6-hours in the NVIDIA Tesla P100 GPU system.

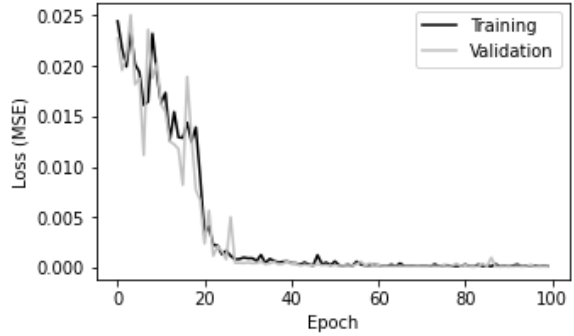

**Figure 5. LSTM training history on 12 cyclic simple shear tests on Nevada sand.**

## RESULTS AND DISCUSSION

Fig. 6 shows the LSTM pore pressure prediction for loose samples ($D_r$ = 41% and 35%) under modulated-up and modulated-down loading conditions. The LSTM model captures the pore water pressure response in laboratory experiments. A hyperbolic tangent activation function restricts the maximum values of predicted $r_u$ to 1.0 to ensure that the excess pore water pressure cannot be larger than the theoretical vertical effective stress. However, the $r_u$ values generated in the laboratory experiment exceed this threshold slightly.

Although the input stress history (modulated up and down) has the same amount of input stress accumulation and amplitudes, the nature of the loading history yields different pore pressure responses. In the modulated-up loading (Fig. 6a), $r_u$ gradually increases after every stress cycle until it reaches 1.0. The amplitude of each cycle exceeds the previous loading peak; hence the soil does not experience any shielding effect. On the other hand, in the modulated-down loading (Fig. 6b), only the first few cycles of loads contribute to $r_u$ build-up. The smaller amplitudes of subsequent cycles only cause a slight increase in $r_u$. The LSTM model captures this shielding effect under different loading conditions. In the modulated-up loading condition (Fig. 6a), where shielding does not occur, $r_u$ reaches 1.0. Whereas in modulated-down loading (Fig. 6b), which shows shielding, $r_u$ does not reach the critical value of 1.0. The LSTM model accurately predicts the $r_u$ response and the required time to liquefaction.

Fig. 7 shows the pore pressure response of a dense sand ($D_r$ = 85%) under modulated-up loading. Similar to the loose sand in modulated-up loading, the pore pressure response $r_u$ gradually increases with each loading cycle. However, the dense sand requires larger stress amplitudes to initiate liquefaction than the loose sand. The LSTM model accounts for the difference in the relative densities by showing a gradual build-up of pore pressure, reaching $r_u$ = 1.0 at the same time as the experiments. The LSTM model can distinguish the effect of density on the pore water pressure response and liquefaction triggering.

Another promising aspect of the LSTM model is its ability to predict the fluctuations in the pore pressure response to cyclic loading, which was not captured by constitutive models such as PM4Sand and PDMY02 (see Fig. 1).

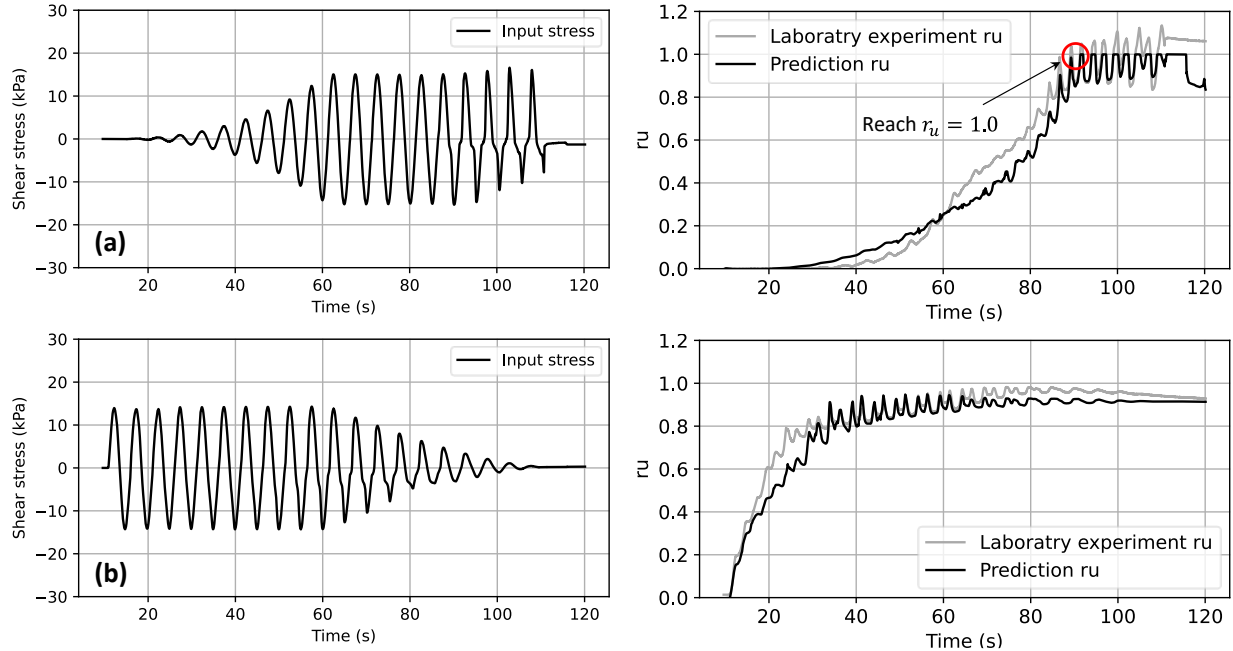

**Figure 6. LSTM prediction of pore pressure response compared with laboratory experiment: (a) Loose sample with $D_r$=41% and modulated-up loading and (b) Loose sample with $D_r$=35% and modulated-down loading.**

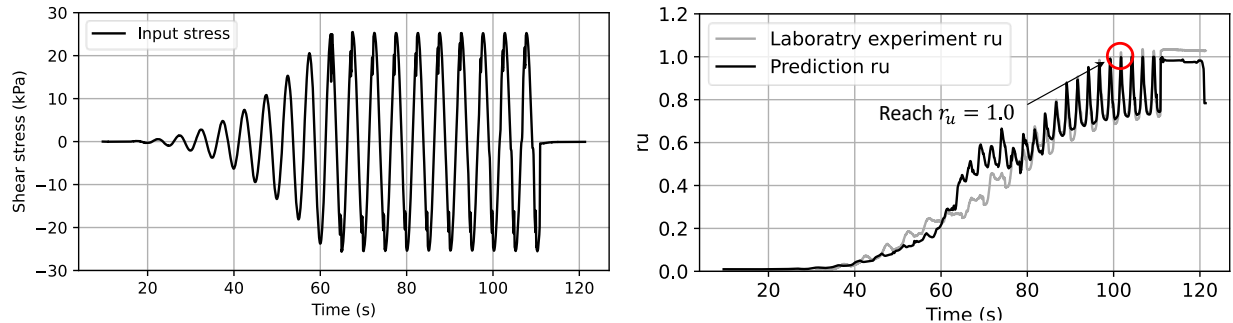

**Figure 7. LSTM prediction of pore pressure response for a dense sample with $D_r$=85% for modulated-up loading**

The LSTM prediction of pore pressure response agrees well with the laboratory results and captures the shielding and density effects. Further studies are needed to capture the pore pressure response of LSTM performance for earthquake excitation. Although we did not include a mechanism in LSTM to evaluate if the current loading is larger than or smaller than the past peak (shielding effect), the model captures this response. However, including such a feature to track if the current stress exceeded prior peak amplitude and using current $r_u$ to predict future pore pressure response is likely to improve the model performance.

# CONCLUSION

Pore water pressure response in liquefiable soils highly depends on shear stress history, such as the shielding effect - where the pore pressure response depends on the loading amplitude compared to the peak prior amplitude. Existing constitutive models fail to capture the shielding effect observed in the cyclic liquefaction experiments. We develop a data-driven machine learning model based on the LSTM neural network to capture sands' liquefaction response under cyclic loading. The LSTM model is trained on 12 laboratory cyclic simple shear tests of sands compacted to different relative densities and loading conditions, i.e., harmonic, modulated-up, and modulated-down loadings. LSTM model features include the relative density and stress history to predict the pore water pressure response. The LSTM model successfully predicts the pore pressure response for three cyclic simple test results considering the shielding and density effects. The model's ability to generalize to earthquake loadings and different ranges of relative densities is not yet proven.

# REFERENCES

Benabou, L. (2021). "Development of Lstm Networks for Predicting Viscoplasticity with Effects of Deformation, Strain Rate, and Temperature History." *Journal of Applied Mechanics*, 88(7), 071008.

Bengio, Y., Simard, P., and Frasconi, P. (1994). "Learning Long-Term Dependencies with Gradient Descent Is Difficult." *IEEE Transactions on Neural Networks*, 5(2), 157-166.

Boulanger, R. W., and Ziotopoulou, K. (2017). *PM4Sand (Version 3.1): A Sand Plasticity Model for Earthquake Engineering Applications*. Center for Geotechnical Modeling, University of California, Davis, CA.

Hochreiter, S., and Schmidhuber, J. (1997). "Long Short-Term Memory." *Neural Computation*, 9(8), 1735-1780.

Ishibashi, I., Sherif, M. A., and Tsuchiya, C. (1977). "Pore-Pressure Rise Mechanism and Soil Liquefaction." *Soils and Foundations*, 17(2), 17-27.

Kwan, W. S., Sideras, S. S., Kramer, S. L., and El Mohtar, C. (2017). "Experimental Database of Cyclic Simple Shear Tests under Transient Loadings." *Earthquake Spectra*, 33(3), 1219-1239.

Sideras, S. S. (2019). *Evolutionary Intensity Measures for More Accurate and Informative Evaluation of Liquefaction Triggering*. Ph.D. dissertation, University of Washington, WA.

Yang, Z., Lu, J., and Elgamal, A. (2008). *Opensees Soil Models and Solid-Fluid Fully Coupled Elements: User's Manual*, Department of Structural Engineering, University of California, San Diego, CA.

Zhang, N., Shen, S., Zhou, A., and Jin, Y. F. (2021). "Application of LSTM Approach for Modelling Stress-Strain Behaviour of Soil." *Applied Soft Computing*, 100, 106959.

Zhang, P., Yin, Z.-Y., Jin, Y.-F., and Ye, G.-L. (2020). "An AI-Based Model for Describing Cyclic Characteristics of Granular Materials." *International Journal for Numerical and Analytical Methods in Geomechanics*, 44(9), 1315-1335.

Zhang, R., Chen, Z., Chen, S., Zheng, J., Büyüköztürk, O., and Sun, H. (2019). "Deep Long Short-Term Memory Networks for Nonlinear Structural Seismic Response Prediction." *Computers & Structures*, 220, 55-68.